# Analysis and Modeling of Social Influence in High Performance Computing Workloads

Shuai Zheng[2], Zon-Yin Shae[1], Xiangliang Zhang[2], Hani Jamjoom[1], and Liana Fong[1]

[1] IBM T. J. Watson Research Center, Hawthorne, NY
[2] King Abdullah University of Science and Technology, Thuwal, Saudi Arabia

**Abstract.** Social influence among users (e.g., collaboration on a project) creates bursty behavior in the underlying high performance computing (HPC) workloads. Using representative HPC and cluster workload logs, this paper identifies, analyzes, and quantifies the level of social influence across HPC users. We show the existence of a social graph that is characterized by a pattern of dominant users and followers. This pattern also follows a power-law distribution, which is consistent with those observed in mainstream social networks. Given its potential impact on HPC workloads prediction and scheduling, we propose a fast-converging, computationally-efficient online learning algorithm for identifying social groups. Extensive evaluation shows that our online algorithm can (1) quickly identify the social relationships by using a small portion of incoming jobs and (2) can efficiently track group evolution over time.

## 1 Introduction

Wide-use and expansion of collaboration technologies (e.g., social networking) are influencing user behavior across all daily activities. Almost completely overlooked, this paper analyzes the effects of *social influence* on high performance computing (HPC) workloads. The intuition is that user collaboration affects the underlying job submission characteristics. For example, students in a class will likely exhibit correlated workload characteristics, especially considering project deadlines, homework, etc.

Discovering the underlying social patterns and dependencies within groups of correlated users—or *communities*, for short—will help improve workload prediction and job scheduling. Our work is akin to those in *community centric web search*, and more recently to Lin *et al.* [7], which discovers the communities based on mutual awareness from observable blogger actions. Unlike existing studies, this paper—to the best of our knowledge—is the first attempt to propose a social-influence-aware method for discovering correlated users and modeling their corresponding workloads in HPC environments.

In an HPC environment, community discovery has several challenges. First, not all the users are regular users of HPC/clusters. Ephemeral users need to be identified and discarded. Second, computing similarities between users is difficult. Since each user submits a different number of jobs to HPC/clusters, measuring the pair wise similarity of users based on their submitted jobs is both complex and unreliable, especially when jobs are described by a complex structured language, e.g., Job Description Language (JDL) [10]. Finally, the community discovery process must be computationally efficient—especially for large-scale workloads—so that it can be used to improve the underlying job scheduling. The challenges outlined above limit the applicability of standard clustering techniques (e.g., double-clustering approach in [12]). In this paper, an efficient method is proposed to identify the *correlated users* in HPC/Cluster workloads.

Following similar analysis of social networks [4], we show that our discovered communities exhibit power-law characteristics. Also, depending on a user's activity, we show that he/she can be categorized as either a *dominant user* or a *follower* to a dominant user. This has profound implications on the importance of dominant users in job scheduling. Basically, identifying dominant users and their followers allows job schedulers to better predict change in future resource demands. To enable effective usage of this new insight, we propose an online learning algorithm that dynamically learns the social characteristics of a given workload. Experimental results show that our online algorithm can efficiently identify stable social groups by observing only a small portion of workload arrivals and can track the group evolution over time.

The remainder of this paper is organized as follows. Section 2 introduces the data sources that we used. We describe our proposed method in Section 3. We then characterize various aspects of the discovered communities in Section 4. Section 5 describes the online learning mechanism. We discuss the related work in Section 6 and conclude the paper in Section 7.

## 2 Data Sources

The first dataset we used is the Grid 5000 traces [2]—a popular HPC workload testbed. Grid 5000 is an experimental grid platform consisting of nine geographically distributed sites across France. Each site comprises of one or more clusters, for a total of 15 clusters. We use the traces recorded by the individual Grid 5000 clusters from the beginning of the Grid 5000 project (during the first half of 2005) to November $10^{th}$, 2006. While there are many useful parameters for each job record in the trace, we extract only `UserID`, `GroupID`, and `SubmittedTime`. In the Grid 5000 trace, there are a total of 10 groups, more than 600 users, and more than 100,000 jobs.

The second dataset that we used is the job trace (i.e., the Logging and Bookkeeping (L&B) files) from the Enabling Grid for ESciencE[1] (EGEE) grid. EGEE currently supports up to 300,000 jobs per day on a $24 \times 7$ basis. Similar to Grid 5000, we extract the submission `timestamp` and `userID` as job parameters. We use two sets of EGEE L&B files: one contains 229,340 jobs submitted by 53 users in 2005; the other contains 347,775 jobs submitted by 74 users in 2007.

[1] http:// www.eu-egee.org/

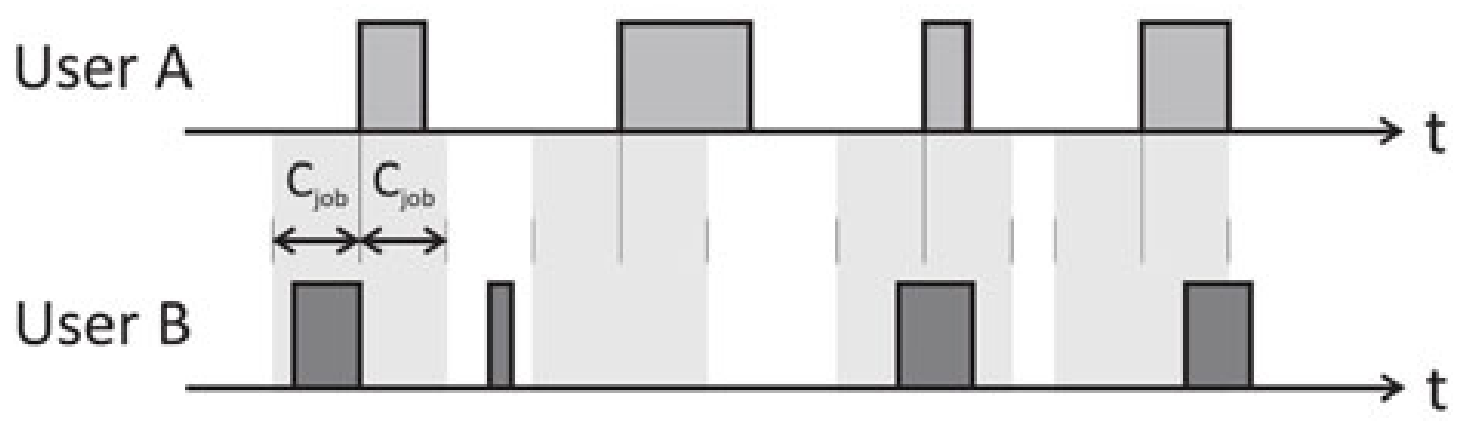

**Fig. 1.** An example of socially-connected jobs. User B has 3/4 (75%) of jobs within a $C_{job}$ before or after a job by User A.

## 3 Social Influence Model

In this section, we define *social influence matrix* (SIM) between users. We focus on studying social relationships based on the submission time of jobs. As an example, consider two users working on a project, which consists of many of jobs. If the two users are working closely on the project (e.g., paper deadline), their job submission times will be close. We refer to the two users and their jobs as being *socially influenced*. In this paper, we use two key features, `UserID` and `SubmittedTime`, to analyze the social influence between users. We do not consider the duration time of jobs.

Take Group 1 in the Grid 5000 dataset as an example. In total, we have 38 different users, each submitting hundreds of jobs. Furthermore, consider two users: User A (submitting 1000 jobs) and User B (submitting 800 jobs). Social influence between these two users is captured by two factors:

- **Socially-connected jobs**: for a job of User B, if the minimum time between the submission time of this job and at least one of the 1000 jobs of User A is small enough (e.g., less than half hour before/after submission time), then we say this job is socially connected to User A. We refer to this minimum time threshold between jobs as $C_{job}$.
- **Socially-connected users**: for the entire 800 jobs of User B, if more than $x\%$ (e.g., $x\% = 50\%$ or 80%) of the jobs are socially-connected jobs to User A, then we say User B is socially connected to User A. We refer to the $x\%$ threshold as $C_{user}$. Furthermore, we define User A as a *dominant user*, and User B as a *follower* of User A.

It should be noted that the social connections are directed relationships. The fact that User A is socially connected to User B does not mean that User B is also socially connected to User A.

Social influence between User A and User B is depicted in Figure 1. In this example, three out of User B's four jobs have at least one submission from User A within the time interval $C_{job}$. We can say that User B has 75% of his/her jobs socially connected with User A, and that he/she is a follower to User A if $C_{user} \leq 75\%$.

Next, we turn our attention to building the SIM. For ease of processing, we sort the 38 users in Group 1 of the Grid 5000 according to their `UserID`. The SIM is a 38 by 38 matrix, noted as $M$. The element $M(i, j)$ of $i$-th row and $j$-th column denotes the corresponding percentage of the jobs of user $i$ that are socially connected to user $j$. We propose Algorithm 1 to calculate the SIM[2].

**Algorithm 1.** Algorithm of Calculating the Social Influence Matrix (SIM)

**Input**: Data set of jobs $D = \{UserID_i,\ JobTime_i\}$,
the $UserID_i$ submitted a job at time $JobTime_i$
$\mathcal{U}$=$\{U_j\}, j = 1...|\mathcal{U}|$, Distinct UserIDs
Criterion $C_{job} = 0.5\ hour, 1\ hour, 6\ hours$
Criterion $C_{user} = 50\%, 80\%$
**Result**: Social Influence Matrix (SIM) $M$
**for** $U_j \in \mathcal{U}$ **do**
  $\mathcal{Y} = \{JobTime_q | \forall q, UserID_q = U_j\}$
    ($JobTime$ of all jobs submitted by $U_j$)
  **for** $U_i \in \mathcal{U}$ **do**
    $\mathcal{X} = \{JobTime_q | \forall q, UserID_q = U_i\}$
      ($JobTime$ of all jobs submitted by $U_i$)
    **for** $k = 1\ to\ |\mathcal{X}|$ **do**
      $d(k) = min(|X(k) - Y(q)|),\ q = 1, ..., |\mathcal{Y}|$
    $M(i, j) = \sum_{k=1}^{|\mathcal{X}|} (d(k) < C_{job}) / |\mathcal{X}|$
    **if** $M(i, j) > C_{user}$ **then**
      user $U_i$ is socially connected to user $U_j$, and
      user $U_i$ is a follower to the dominant user $U_j$

Table 1 shows the SIM of Group 1 when $C_{job}$ is one hour. We give SIM of the first five users. In the first column, the first element is 1, which means that 100% of User 1's jobs are socially connected to himself/herself. Obviously, all diagonal values of the matrix are 1. The second element in the first column is 0, which means that none of the jobs of second user are socially connected to the first user.

As described earlier, social connections between users are influenced by the threshold value $C_{user}$—the minimum percentage of socially-connected jobs. If we set the criterion $C_{user}$ to 80%, the *followers* to User $i$ are the ones who have values larger than 0.8 in $i$-th column. We show the number of *followers* in Table 2 for $C_{user}$=80% and $C_{user}$=50%. As expected, when $C_{user}$ is decreased, we have an increase in the number of followers. For example, User 2 has 5 followers when $C_{user}$ is reduced to 50% from 80%. In the following section, we will plot the results like Table 2 for the two Grid 5000 datasets and the two EGEE datasets to explore their number of followers distribution.

[2] For a large set of users, we have two solutions to handle them in practice. First, we can parallelize Algorithm 1 to efficiently calculate SIM $M$, because the calculations of elements $M_{ij}$ are independent. Second, we can use the later proposed Algorithm 2 to incrementally update $M$ online.

**Table 1.** One-Hour Social Influence Matrix (SIM)

| | User 1 | User 2 | User 3 | User 4 | User 5 | ... |
|---|---|---|---|---|---|---|
| User 1 | 1 | 0 | 0 | 0 | 0 | ... |
| User 2 | 0 | 1 | 0 | 0.029 | 0.084 | ... |
| User 3 | 0 | 0 | 1 | 0 | 0 | ... |
| User 4 | 0 | 0.131 | 0 | 1 | 0.239 | ... |
| User 5 | 0 | 0.048 | 0 | 0.056 | 1 | ... |
| ... | ... | ... | ... | ... | ... | ... |

**Table 2.** Number of Followers. A value of 1 means that there is only one follower to the corresponding user, or this user is only socially connected to himself/herself.

| $C_{job}$ $C_{user}$ | User 1 | User 2 | User 3 | User 4 | User 5 | ... |
|---|---|---|---|---|---|---|
| 1 hr 80% | 1 | 3 | 1 | 2 | 1 | ... |
| 1 hr 50% | 1 | 5 | 1 | 2 | 2 | ... |

## 4 Analysis of Social Influence

In this section, we analyze Grid 5000 and EGEE datasets to discover socially-connected users. As expected, the values for $C_{job}$ and $C_{user}$ will impact the resulting analysis. We present our results for a number of value combinations for both parameters. One challenge was choosing a reasonable parameter range for $C_{job}$. In particular, $C_{job}$ should be set as small as possible to capture true dependencies. To reason about value for $C_{job}$, we plot the Cumulative Distribution Function (CDF) of the jobs' interarrival time for the four datasets in Figure 2. As the figure shows, the interarrival times vary significantly over multiple orders of magnitude. Picking a small value of $C_{job}$ will unnecessarily filter out longer-range (i.e., minutes rather than seconds) dependencies, which are common in human interactions. Figure 2 indicates that—on average—the probability that at least one job will be submitted within 30 minutes for all the groups under study is larger than 95%. Thus, we decide to use $C_{job}$ values in the range from 30 minutes to 6 hours in our investigation.

### 4.1 Community Extraction from HPC Workloads

Figure 3 (a) shows the number of followers for each user identified from Group 1 of Grid 5000 trace with criterion $C_{user}$=50%. When $C_{user}$=50% and $C_{job}$=6 hours, User 1, 2, and 3 have 13, 11, and 10 followers, respectively. When $C_{user}$ =50% and $C_{job}$=1 or 0.5 hours, the number of followers for every user decreases as expected. Figure 3 (b) shows the number of followers for each user in Group 1 with criterion $C_{user}$=80%. From Figure 3 (a) and (b), we can see that all users

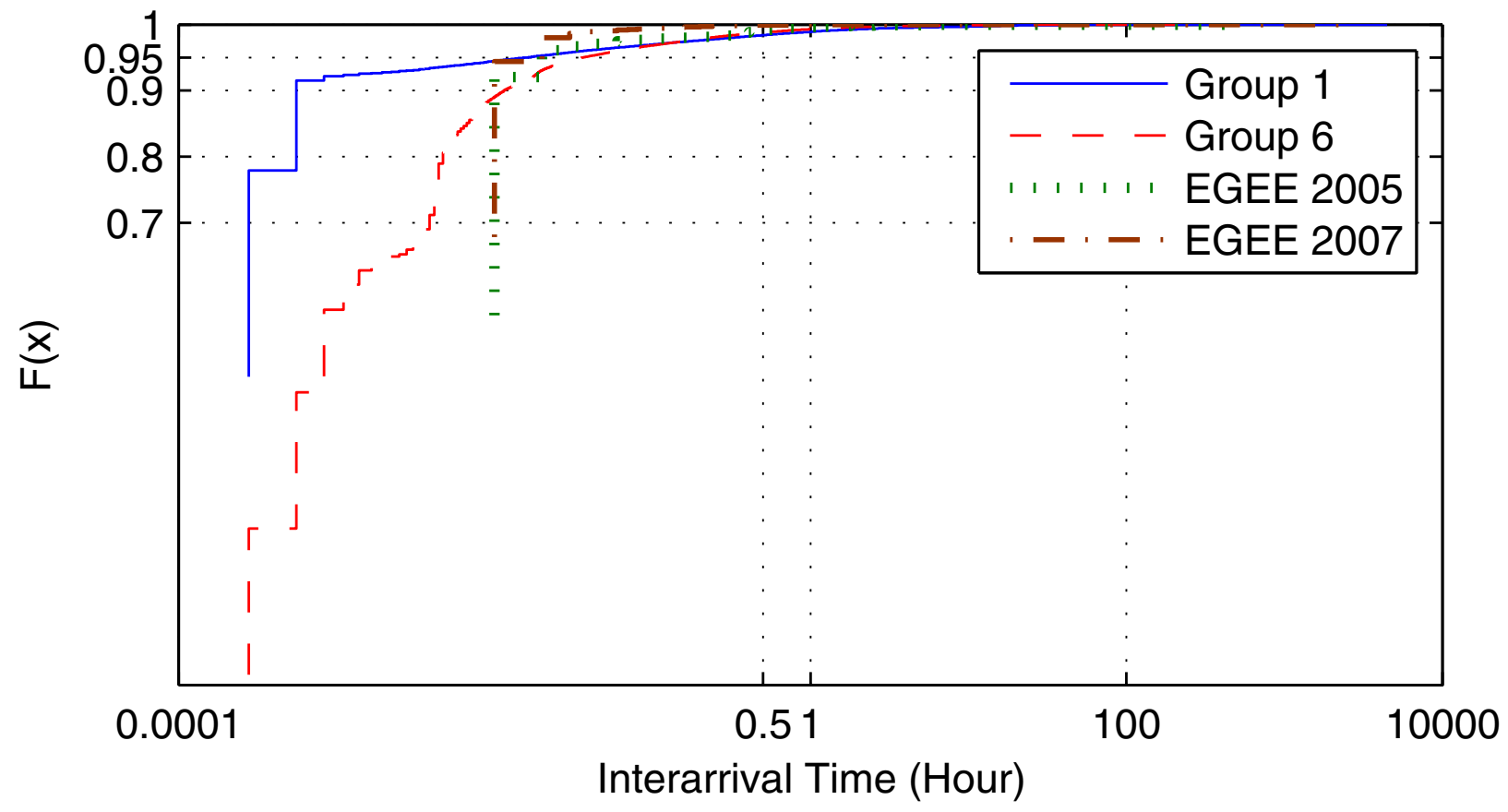

**Fig. 2.** CDF of jobs' interarrival time for Group 1, Group 6, EGEE 2005 and EGEE 2007

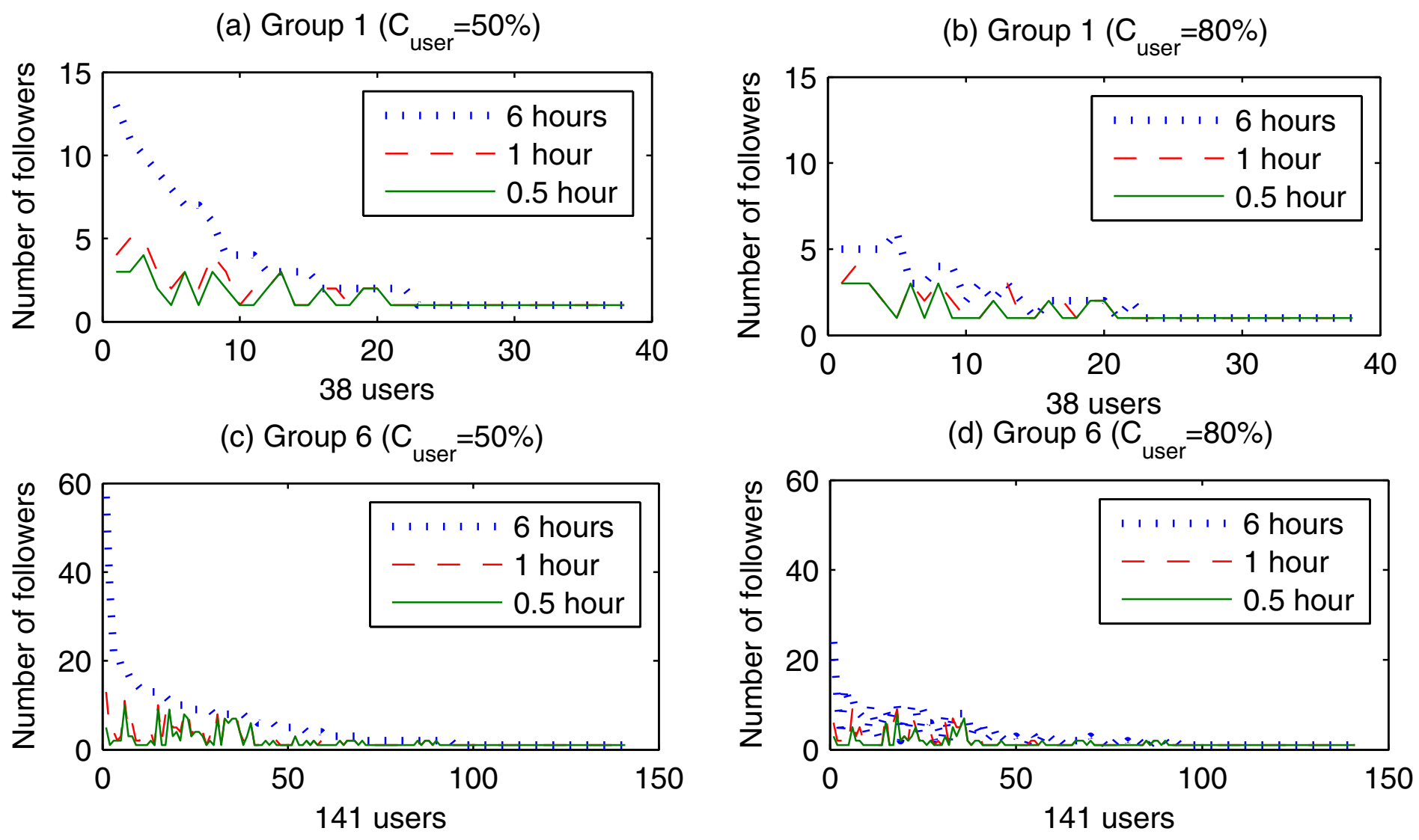

**Fig. 3.** Social groups discovered in Group 1 and Group 6 of Grid 5000 traces showing the number of followers to each dominant user with different criteria of $C_{user} = 50\%$, 80% and $C_{job} = 6\ hours$, $1\ hour$, $0.5\ hour$

in Group 1 consistently have followers. Similarly, Figure 3 (c) and (d) show that nearly 60% of all users in Group 6 consistently have followers. Figure 4 shows that around 45% of all users in EGEE 2005 consistently have followers, and around 55% of all users in EGEE 2007 consistently have followers.

## 4.2 Power-Law Distribution of Discovered Communities

A common property of many large networks is that the vertex connectivities follow a scale-free power-law distribution [4]. We would like to investigate if such a power-law property exists among HPC users. In our study, each vertex is a person. The connectivity among vertices captures the interactions among users.

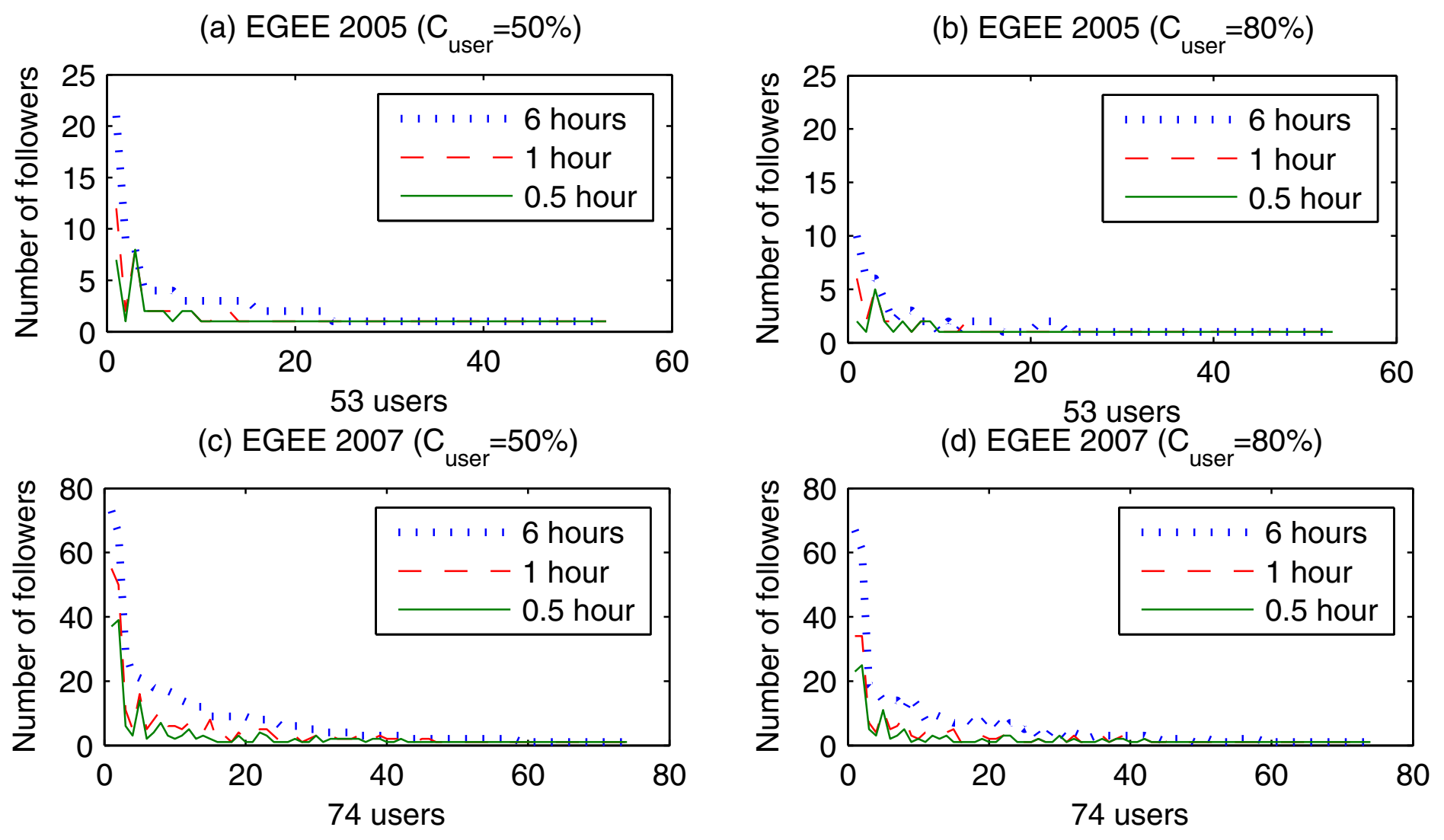

**Fig. 4.** Social groups discovered in EGEE 2005 and EGEE 2007 traces showing the number of followers to each dominant user with different criteria of $C_{user} = 50\%$, 80% and $C_{job} = 6\ hours$, 1 *hour* and 0.5 *hour*

Let $P(k)$ denote the probability that a user has a number of followers $k$, where $k$ is a positive integer number. To study if our discovered social groups have the same property as the common networks, we investigate whether the number of followers $k$ of each dominant user has the power-law distribution: $P(k)$=$a \times k^b$. Figure 5 shows the power-law distribution of the number of followers identified from Group 1 and 6 of Grid 5000, as well as from EGEE 2005 and EGEE 2007. The power-law distributions of other groups exhibit similar characteristics (unless the group size is very small); they are not shown for space consideration. All social followers are discovered with socially-connected criterion $C_{job} = 0.5\ hour$ and $C_{user} = 50\%$. From Figure 5, we can see that the number of followers fits very well the power-law distribution with different parameters $a$ and $b$.

## 5 Design of Online Learning Mechanism

Our earlier analysis used an offline mechanism to identify social influence in HPC workloads. For our analysis to be consumable by HPC job schedulers and resource managers, a real-time (online) mechanism is needed. Algorithm 2 shows the proposed mechanism for computing the social influence matrix (SIM) on the fly while jobs are arriving. Suppose that at time step, $t$, a user submits a job. We have $(UserID^t, JobTime^t)$ as a sample in our streaming workload, where time step $t \geq 1$ is an integer[3]. Given the threshold $C_{job}$ of socially-connected jobs, we maintain a $Un^t \times Un^t$ matrix $M^t$, where $Un^t$ is the number of unique users until $t$. Each element of $M^t$ is a 2-tuple object $M^t_{ij} = \{R^t_{ij}, C^t_i\}$, where $C^t_i$ is the number of jobs submitted by user $UserID = U_i$ until time $t$, and $R^t_{ij}$ is the number of jobs of $UserID = U_i$ that are socially connected to $UserID = U_j$ until $t$. The threshold $C_{job}$ and $C_{user}$ are set to half hour and 50% respectively by taking experience from the offline mechanism for enabling further comparison.

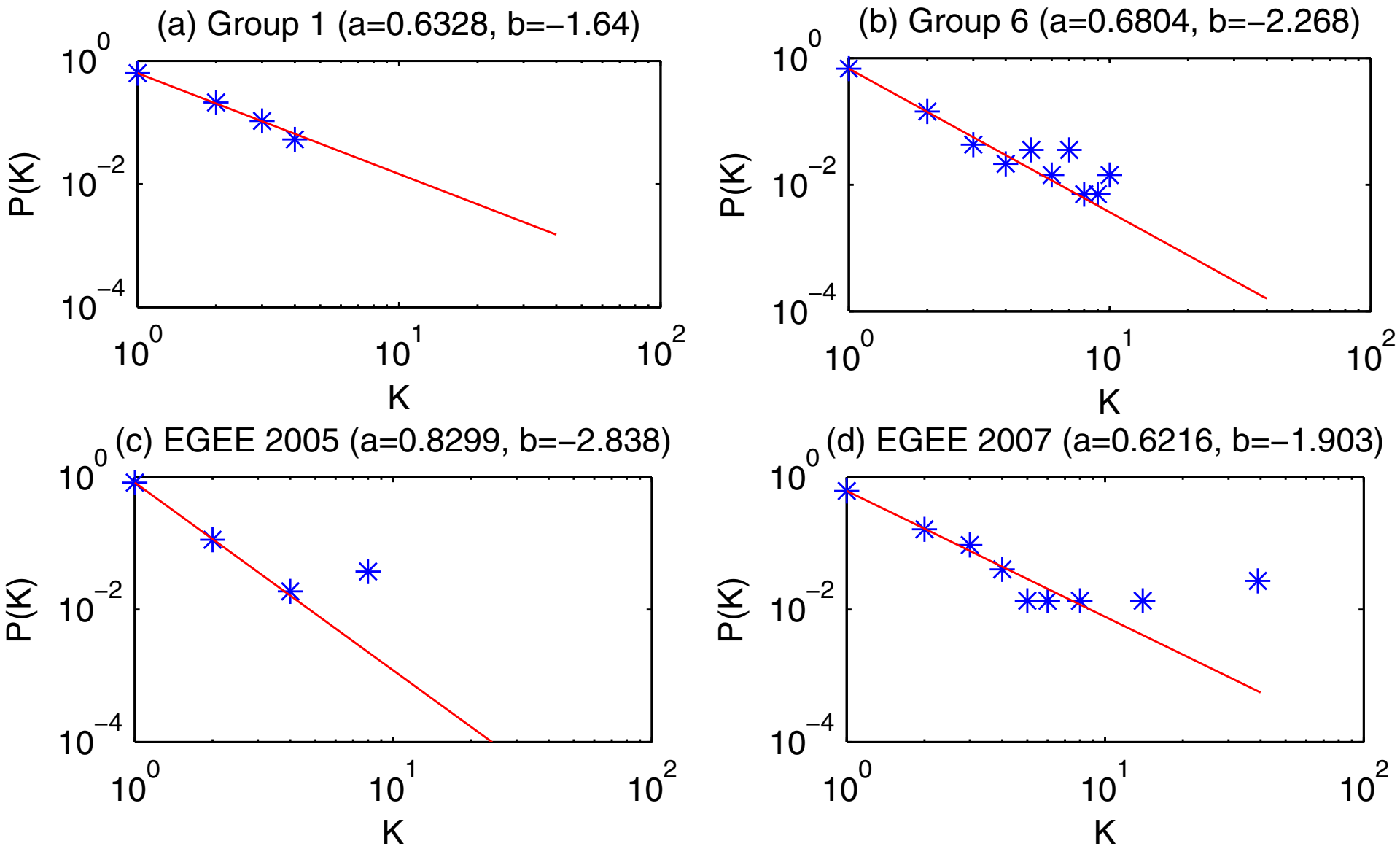

**Fig. 5.** The power-law distribution of the number of users $k$ following each dominant user identified in Group 1 and Group 6 of Grid 5000, EGEE 2005 and EGEE 2007

We use MATLAB to implement and simulate Algorithm 2. Unlike Algorithm 1, we use a sliding time window along the workload flow. This window only includes past job submissions within the interval of $C_{job}$. Note, this is different to the method described in Algorithm 1 for checking the socially-connected jobs. In Algorithm 1, we consider the absolute value of the time difference, which includes both sides of the current time point on the time axis (as shown in Figure 1).

We use *cosine similarity* [13] to measure the difference between the distribution of followers as obtained by Algorithms 1 and 2. We set $C_{job}$=0.5 hour and $C_{user}$=50%. Figure 6 shows the cosine similarity between online results as compared to its offline counterpart as a function of the percentage of observed jobs (in chronological order). The cosine similarity increases towards the value 1 as additional jobs are observed. When all jobs have been observed (at the 100% value on the x-axis), the online and offline algorithm produce the same results (cosine similarity = 1).

A very promising characteristic of the online algorithm is its ability to track group evolution over time. For example, Figure 6 shows how Group 6 and EGEE

[3] Time step $t$ is used to order the streaming jobs by arrival time. If $t1 < t2$, $UserID^{t1}$ submitted a job earlier than $UserID^{t2}$, i.e., $JobTime^{t1} < JobTime^{t2}$.

**Algorithm 2.** Online Algorithm of Calculating Socially Influence Matrix (SIM)

**Input**: Streaming jobs $S = \{UserID^t,\ JobTime^t\}$,
the $UserID^t$ submitted a job at time $JobTime^t$
Criterion $C_{job} = 0.5\ hour$
Criterion $C_{user} = 50\%$
**Result**: Socially Influence Matrix $M^t$, $M^t_{ij} = \{R^t_{ij}, C^t_i\}$
where $R^t_{ij}$ is the number of $U_i$'s jobs socially connected to $U_j$ until $t$, and $C^t_i$ is the number of jobs submitted by user $U_i$ until $t$
**Initialization:**
Distinct users set $\mathcal{U} = \{\}$, and the number of distinct users $Un^t = 0$,
$M^t_{ij} = \{R^t_{ij} = 0, C^t_i = 0\}$
**Maintain** $M^t$
**for** $t = 1$ *to* ... **do**
  **if** $UserID^t \notin \mathcal{U}$ **then**
    $\mathcal{U} = \mathcal{U} \cup UserID^t$
    $Un^t = Un^{t-1} + 1$
  $i \leftarrow UserID^t = U_i$
  $C^t_i = C^{t-1}_i + 1$
  $\mathcal{X} = \{UserID^q\ |\forall q, JobTime^q >= JobTime^t - C_{job}\}$
  a set of $UserID$ whose jobs arrived within a time-window $C_{job}$
  **for** $U_j \in \mathcal{X}$ **do**
    $R^t_{ij} = R^{t-1}_{ij} + 1$
    $R^t_{ji} = R^{t-1}_{ji} + |X^{j,t'}|$
    where $|X^{j,t'}|$ is the number of $U_j$ in time window of $[max\{t - C_{job}, t' + C_{job}\}, t]$ and $t'$ is the last time when $U_i$ appeared
  **Socially-connected users**
  **for** $j = 1$ *to* $Un^t$ **do**
    **if** $R^t_{ij}/C^t_i > C_{user}\ in\ M^t_{ij}$ **then**
      user $U_i$ is socially connected to user $U_j$, and
      user $U_i$ is a follower to the dominant user $U_j$

2005 have sudden increase in the number of discovered social groups after processing 50% and 80% of all job flows, respectively. In contrast, Group 1 and EGEE 2007 have stable social groups after processing 5% of all jobs, with little change beyond that point.

The discovered changes in social groups reflect variations in HPC system usage. In order to verify the changes in social relationships, we investigate the number of distinct users as a function of job arrivals (Figure 7). Comparing Figures 7 and 6, we find that the curves move in tandem (as a function of job arrival). For example, Group 6 and EGEE 2005 have more new users after observing 50% and 80% of all job flows. The number of users in Group 1 and EGEE 2007 grow slowly after 5% of all jobs.

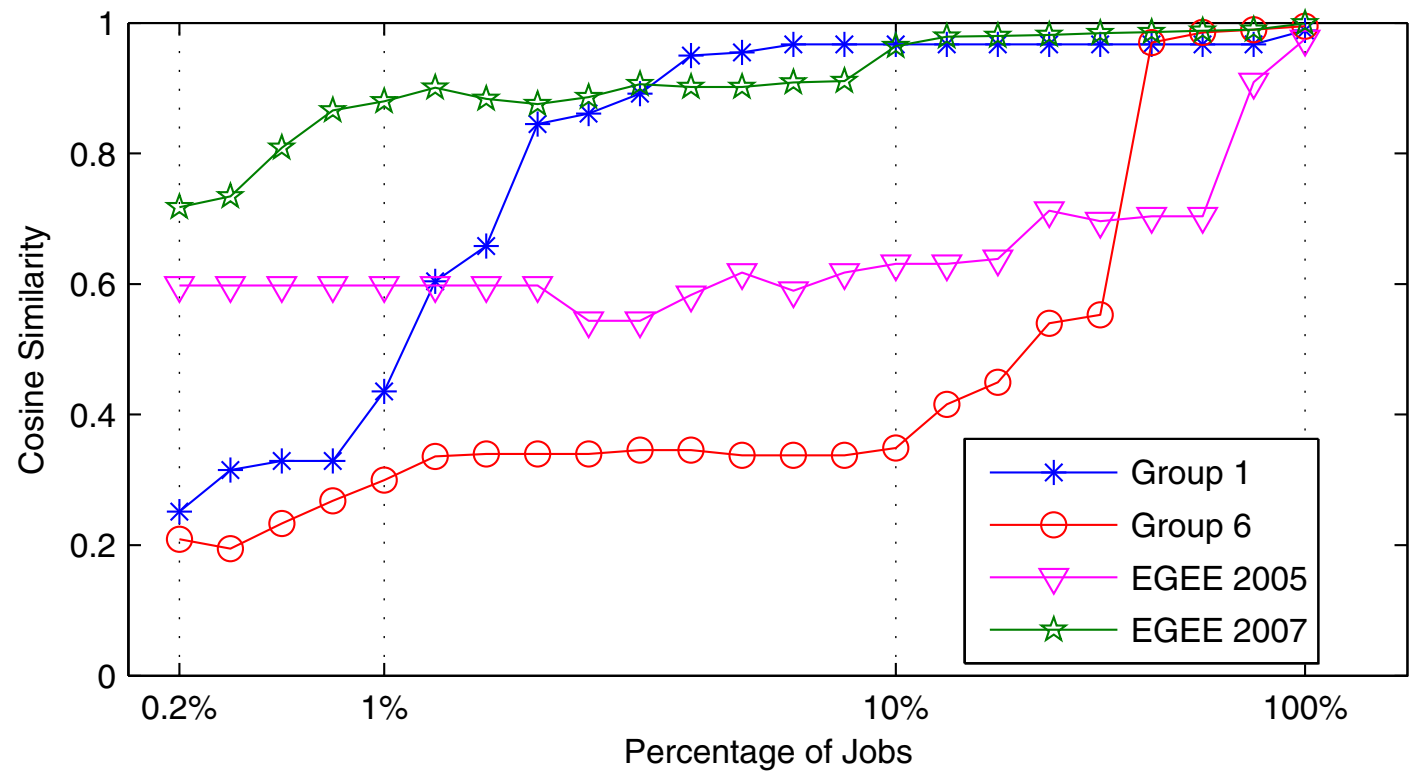

**Fig. 6. Online** Learning Convergence for Group 1, Group 6, EGEE 2005 and 2007

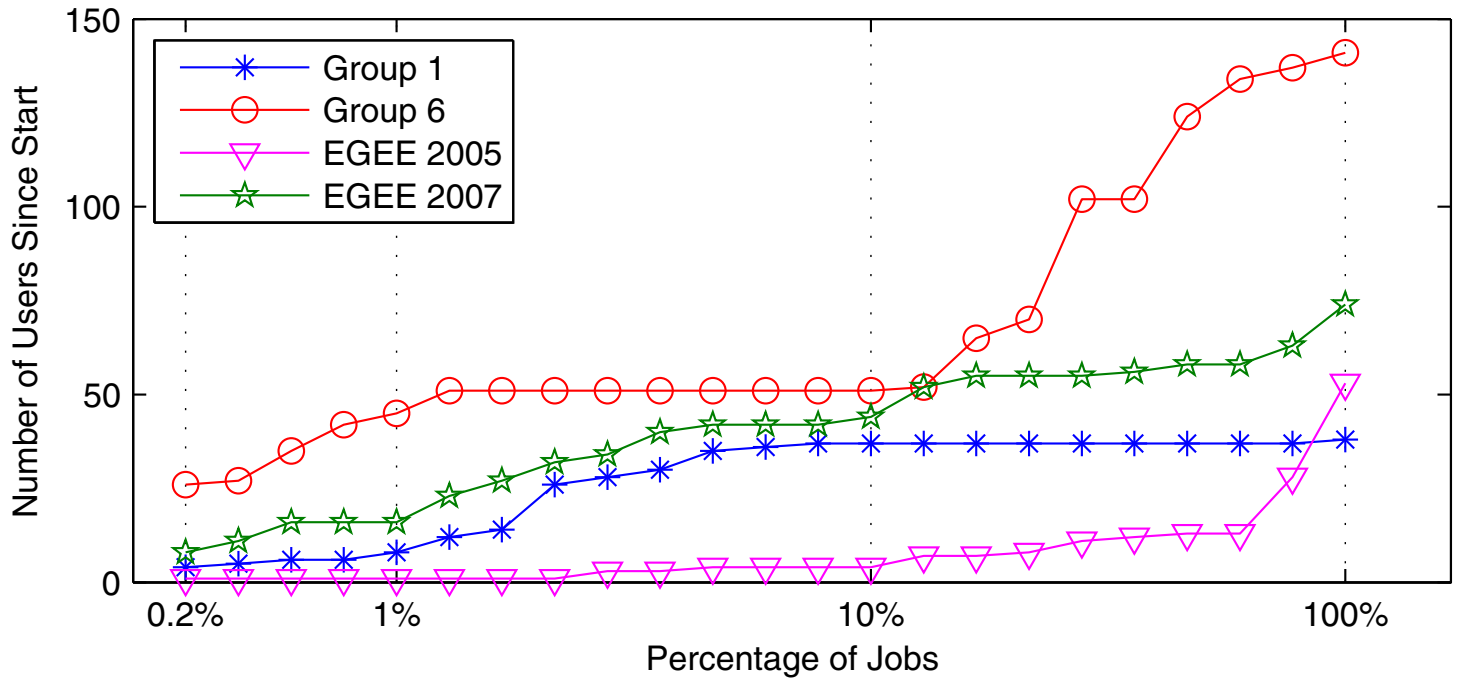

**Fig. 7. Online** Users Count for Group 1, Group 6, EGEE 2005 and EGEE 2007

We are still interested in how fast the similarity can reach a stable state. Figure 6 shows that our online algorithm can converge and reach a stable state quickly in real time. For example in Group 1, we see that the user population (number of distinct users as shown in Figure 7) is stable within 1% of job flows, while the cosine similarity (in Figure 6) reaches the stable state within about 0.4% of all jobs. When the user population of Group 1 changes dramatically from 1% to 7%, Figure 6 shows that the cosine similarity follows the changes quickly and reaches a stable state closing to 1.

## 6 Related Work

Many generally available workload traces [3,2,1] have been used to study design alternatives for resource scheduling algorithms, resource capacity planning in both grid cluster and cloud environments, building performance modeling, etc. Most of these studies use—as input—the arrival pattern of individual jobs and their resource requirements [8]. Only recently, Iosup *et al.* [5] presented their first investigation of the grouping of jobs by looking at their submission patterns and their impact on computing resource consumption. More recently, Ostermann *et al.* [9] studied job flows focusing on how sub-jobs are submitted in parallel or in sequence [9]. Unlike existing work, we investigate job submission patterns from the perspective of social connections (e.g., group association, virtual organization, etc.). Especially in cloud environments, the ability to predict future demands is critical to managing the underlying computing resources.

Community extraction has been commonly studied as a graph problem. A graph is constructed by taking users/persons as nodes and connecting two nodes if they are correlated in some activities. A community discovered in a graph is a subgraph including a group of nodes and their edges, where the nodes have high similarity with each other. Two approaches have been used to identify the subgraphs. One is based on a clustering method, weighted kernel k-means, which groups together the nodes that are similar to each other by measuring a type of random walk distance [14]. The other approach uses graph partitioning algorithm (or called spectral clustering) to cut the graph into a set of subgraphs by optimizing the cost of cutting edges under the normalized cut criterion [11].

The most relevant work to our paper is the Mixed User Group Model (MUGM) proposed by Song *et al.* [12]. MUGM forms the groups of users through characterizing each user by job clusters, which were obtained by using CLARA (also known as k-medoids) clustering method [6] on the whole jobs. There are two difficulties in applying MUGM on analyzing HPC workloads. First, using clustering method to group jobs requires the computation of job similarities, which is difficult for jobs described by mixture of features. Second, representing users based on job clusters cannot be easily adapted as an online technique, because job clusters need to be computed on the whole data before analyzing the user groups. Our approach identifies social groups by measuring their tasks' submission behavior. Furthermore, our approach does not need the computation of job similarities and can be efficiently used in an online fashion.

## 7 Conclusions and Future Work

This paper identified and validated the existence of social influence on HPC workloads. We suspect that this influence stems from how HPC applications are developed and run. Given its potential importance on job scheduling and resource management, we proposed a method to discover *socially-connected users* based on measuring the proportion of *socially-connected jobs* they have. We showed the existence of a social graph characterized by a pattern of dominant users and followers. We applied this proposed method to traces of Grid 5000 and EGEE. Both in the Grid 5000 and the EGEE workloads, we consistently found that around half of the users had *followers* irrespective of how the thresholds $C_{job}$ and $C_{user}$ were set. Additionally, the corresponding social graph followed a power-law distribution, which is consistent with mainstream social networks. We developed both an offline and a fast-converging online algorithm to implement our proposed method. The online version was shown to require a small number of arrived jobs to discover the social groups and be able to track the group evolution over time.

Identifying dominant users and their followers may have profound implication on the prediction of workload and, consequently, on resource demand patterns. Thus, our future work will focus on quantifying social characteristics of the discovered user connections and using this quantitative information to improve the prediction of workload arrival patterns and resource demands. The prediction will enable the development of new algorithms in job scheduling, resource utilization, and resource capacity planning.